# Imaging electrostatically confined Dirac fermions in graphene quantum dots

Juwon Lee[1]*, Dillon Wong[1]*, Jairo Velasco Jr.[1,2†], Joaquin F. Rodriguez-Nieva[3], Salman Kahn[1], Hsin-Zon Tsai[1], Takashi Taniguchi[4], Kenji Watanabe[4], Alex Zettl[1,5,6], Feng Wang[1,5,6], Leonid S. Levitov[3] and Michael F. Crommie[1,5,6†]

[1]*Department of Physics, University of California, Berkeley, California 94720, USA*
[2]*Department of Physics, University of California, Santa Cruz, California 95064, USA*
[3]*Department of Physics, Massachusetts Institute of Technology, 77 Massachusetts Ave, Cambridge, Massachusetts 02139, USA*
[4]*National Institute for Materials Science, 1-1 Namiki, Tsukuba, 305-0044, Japan*
[5]*Materials Sciences Division, Lawrence Berkeley National Laboratory, Berkeley, California 94720, USA*
[6]*Kavli Energy NanoSciences Institute at the University of California, Berkeley and the Lawrence Berkeley National Laboratory, Berkeley, California 94720, USA*

*\*These authors contributed equally to this work.*
*†crommie@berkeley.edu, jvelasc5@ucsc.edu*

**Electrostatic confinement of charge carriers in graphene is governed by Klein tunneling, a relativistic quantum process in which particle-hole transmutation leads to unusual anisotropic transmission at pn junction boundaries[1-5]. Reflection and transmission at these novel potential barriers should affect the quantum interference of electronic wavefunctions near these boundaries. Here we report the use of scanning tunneling microscopy (STM) to map the electronic structure of Dirac fermions confined by circular graphene pn junctions. These effective quantum dots were fabricated using a new technique involving local manipulation of defect charge within the insulating substrate beneath a graphene monolayer[6]. Inside such graphene quantum dots we observe energy levels corresponding to quasi-bound states and we spatially visualize the quantum interference patterns of confined electrons. Dirac fermions outside these quantum dots**

**exhibit Friedel oscillation-like behavior. Bolstered with a theoretical model describing relativistic particles in a harmonic oscillator potential, our findings yield new insight into the spatial behavior of electrostatically confined Dirac fermions.**

Electrons within graphene have previously been confined by lithographically patterned structures[7-10], graphene edges[11], and chemically synthesized graphene islands[12-15], but these systems are either too contaminated for direct wavefunction visualization or utilize metallic substrates that prevent electrostatic gating. Electron confinement in graphene has also been induced through magnetic fields[16] and supercritical impurities[17], but these methods are unwieldy for many technological applications. An alternative approach for confining electrons in graphene relies on using electrostatic potentials. However, this is notoriously difficult because Klein tunneling causes electric potentials to become transparent to massless Dirac fermions at normal incidence[1-5]. Nevertheless, it has been theoretically predicted that a circular graphene pn junction can localize Dirac electrons and form quasi-bound quantum dot states[18-23]. A recent tunneling spectroscopy experiment by Zhao *et al.*[24] revealed signatures of electron confinement induced by the electrostatic potential created by a charged scanning tunneling microscope (STM) tip. However, since the confining potential moves with the STM tip, this method neither allows spatial imaging of the resulting confined modes nor allows further patterning control of the confinement potential.

Here we employ a new patterning technique that allows the creation of stationary circular pn junctions in a graphene layer on top of hexagonal boron nitride (BN). Figure 1a illustrates how stationary circular graphene pn junctions are created. We start with a graphene/BN heterostructure resting on an $SiO_2$/Si substrate. The doped Si substrate acts as a global backgate while the BN layer provides a tunable local embedded gate after being treated by a voltage pulse from an STM tip[6]. To create this embedded gate the STM tip is first retracted approximately 2 nm above the graphene surface and a voltage pulse of $V_s = 5$ V is then applied to the STM tip while simultaneously holding the backgate voltage to $V_g = 40$ V. The voltage pulse ionizes defects in the BN region directly underneath the tip[25] and the released charge migrates through the BN. This leads to a local space-charge buildup in the BN that effectively screens the backgate and functions as a negatively charged local embedded gate[6] (using the opposite polarity gate voltage during this process leads to an opposite polarity space charge). Adjusting $V_g$ afterwards allows us to tune the overall doping level so that the graphene is n-doped globally but p-doped inside a circle centered below the location where the tip pulse occurred (it is also possible to create opposite polarity pn junctions). As shown schematically in Fig. 1b, the STM tip can then be moved to different locations to probe the electronic structure of the resulting stationary circular pn junction.

To confirm that this procedure results in a circular pn junction, we measured STM differential conductance ($dI/dV_s$) as a function of sample bias ($V_s$) on a grid of points covering the graphene area near a tip pulse. The Dirac point energy, $E_D$, was identified at every pixel, allowing us to map the charge carrier density, $n$, through the relation $n(x,y) = -\frac{\text{sgn}(E_D)\, E_D^2}{\pi (\hbar v_F)^2}$, where $v_F = 1.1 \times 10^6$ m/s is the graphene Fermi velocity[26,27]. Figure 1c shows the resulting $n(x,y)$ for a tip pulse centered in the top right corner. The interior blue region exhibits positive charge density (p-type) while the red region outside has negative charge density (n-type).

In order to spatially map the local electronic properties of such circular pn junctions, we examined a rectangular sector near a pn junction as indicated in Fig. 2a. Figure 2b shows a topographic image of the clean graphene surface in this region. A 2.8 nm moiré pattern is visible[28,29] and the region is seen to be free of adsorbates. A $dI/dV_s$ map of the same region (Fig.

2c) reflects changes in the local density of states (LDOS) caused by the spatially varying charge density distribution. Since the pn junction center is stationary, we are able to move the STM tip to different locations inside and outside the pn junction in order to spatially resolve the resulting electronic states. Figures 2d-g show $d^2I/dV_s^2(V_g,V_s)$ plots at four different locations as denoted in Fig. 2c. We plot the derivative of $dI/dV_s$ with respect to $V_s$ in order to accentuate the most salient features, which are quasi-periodic resonances that disperse to lower energies with increasing $V_g$ (see Supplementary Information for $dI/dV_s$ sweeps before differentiation). The energies of the observed resonances are seen to evolve as $\varepsilon \propto \sqrt{|V_g - V_{CNP}|}$ + constant, where $V_{CNP}$ is the local charge neutrality point, as expected for graphene's relativistic band structure. We see that the energy spacing between resonances ($\Delta\varepsilon$) decreases as we move away from the pn junction center until the resonances disappear outside. For example, $\Delta\varepsilon$ is $29 \pm 2$ mV at the center, $16 \pm 2$ mV 50 nm from the center, and $13 \pm 2$ mV 100 nm from the center (for $V_g = 32$ V). A similar trend is also observed for pn junctions that are n-doped in the center and p-doped outside (see Supplementary Information).

We have imaged these electronic states both inside and outside of circular pn junctions. The $dI/dV_s$ maps in Figs 3a and 3b show eigenstate distributions mapped at two different energies within the same section of a circular pn junction similar to the boxed region of Fig. 2a (but with opposite heterojunction polarity). Circular quantum interference patterns resulting from confined Dirac fermions are clearly observed within the junction boundary, as well as scattering states exterior to the boundary. The junction boundary is demarcated by a dark band (low $dI/dV_s$) in the middle of each $dI/dV_s$ map (and further marked by a dashed line). The two eigenstate distributions are clearly different since one has a node at the origin while the other exhibits a central anti-node. Figure 4a shows a more complete mapping of the energy-dependent eigenstates (within a pn junction of the same polarity as Fig. 2a) along a line extending from the center (left edge) to a point outside of the pn junction (right edge) at a gate voltage of $V_g = 32V$. The data are plotted as $d^2I/dV_s^2(r,V_s)$ (where $r$ is the radial distance from the center) to accentuate the striking oscillatory features (see Supplementary Information for $dI/dV_s(r,V_s)$ prior to differentiation). The energy level structure and interior nodal patterns are clearly evident.

Our observations can be explained by considering the behavior of massless Dirac fermions in response to a circular electrostatic potential. Due to Klein tunneling, a graphene pn junction perfectly transmits quasiparticles at normal incidence to the boundary but reflects them at larger angles of incidence[1,4,5]. In a potential well with circular symmetry electrons with high angular momenta are obliquely incident on the barrier and are internally reflected, thus leading to particle confinement and the formation of quasi-bound quantum dot states[19-24]. As angular momentum is increased, electrons are repelled from the center of the potential by the centrifugal barrier, leading to an increase in the number of $dI/dV_s$ resonances that should be observable in spectroscopy measured away from the center[30]. This is consistent with our observation that the energy spacing between resonances ($\Delta\varepsilon$) at the center (Fig. 2d) is approximately double the energy spacing at a point 100 nm away (Fig. 2f). Scattered quasiparticles external to the potential boundary contribute to Friedel oscillations that radiate outwards, as seen in Fig. 3. A circular graphene pn junction with an n-doped interior thus acts as a quantum dot for electron-like carriers and a quantum antidot for hole-like carriers (as in Fig. 3), and the reverse is true for pn junctions of opposite polarity (as in Figs 2 and 4).

This qualitative picture can be quantitatively tested by comparing the experimental results to a model based on the two-dimensional massless Dirac Hamiltonian, $H = -i\hbar v_F \boldsymbol{\sigma} \cdot \nabla_{\boldsymbol{r}} + U(\boldsymbol{r})$, where $U(\boldsymbol{r}) = -\kappa r^2$ is a quadratic potential and $\boldsymbol{\sigma} = (\sigma_x, \sigma_y)$ are the Pauli matrices.

The strength of the potential, $\kappa = 6 \times 10^{-3}$ meV/nm$^2$, was extracted from measurements of the spatially-dependent Dirac point energy (see Supplementary Information). By solving the Dirac equation under these conditions we are able to determine the allowed eigenstates for Dirac fermions in this confinement potential, as well as their spatial distribution. Figure 4b shows the results of our calculations in a plot of $\partial\text{LDOS}/\partial\varepsilon$, the energy derivative of the Dirac fermion LDOS ($\partial\text{LDOS}/\partial\varepsilon$ corresponds to the experimental quantity $d^2I/dV_s^2$). This theoretical eigenstate distribution (Fig. 4b) closely resembles the experimental eigenstate distribution (Fig. 4a). Both have a characteristic parabolic envelope due to the confinement potential as well as a complex set of interior nodal patterns. The energy spacing of the modes close to the center ($r$ = 0 nm) is seen to be larger than for the modes further out since only the lowest angular momentum states reach the origin. The characteristic energy spacing seen experimentally is in good agreement with the characteristic energy scale $\varepsilon^* = (\hbar^2 v_F^2 \kappa)^{1/3} \approx 15$ meV from the theoretical model.

Further insight into the nature of these confined modes can be gained by directly comparing constant-energy experimental $dI/dV_s$ line cuts (Fig. 4c) to the simulated quantum dot wavefunctions (Fig. 4d). Here it is useful to label the confined states by a radial quantum number $n = 0, 1, 2, ...$ and an azimuthal quantum number $m = \pm\frac{1}{2}, \pm\frac{3}{2}, ...$, i.e. $H\Psi_{n,m} = \varepsilon_{n,m}\Psi_{n,m}$. In order to understand the experimentally observed behavior we note two important features of the eigenstates $\Psi_{n,m}$. First, while each probability distribution $|\Psi_{n,m}|^2$ possesses $n + 1$ maxima, most of the weight is centered in the first maximum which is pushed further from the center for larger values of $|m|$ (Fig. 4d). Thus we are able to attribute each experimental $dI/dV_s$ peak in Fig. 4c to a different $(n, m)$ state since each eigenstate contributes most of its spectral weight to a single energy and radial position. Second, for massless Dirac fermions confined by a quadratic potential there happens to be strong energy alignment of the states $\varepsilon_{n,m}, \varepsilon_{n-1,m+2}, ...$ at low quantum numbers, indicating a near accidental degeneracy. This explains the origin of the multiple peaks observed in many of the d$I$/d$V_s$ line scans of Fig. 4c. These peaks have a spatial extent that is similar to the characteristic length scale $r^* = (\hbar v_F/\kappa)^{1/3} \approx 50$ nm exhibited by Dirac equation solutions.

In conclusion, we have spatially mapped the electronic structure inside and outside of highly tunable quantum dots formed by circular graphene pn junctions. Within these quantum dots we observe the energy levels and nodal patterns that result from the electrostatic confinement of Dirac fermions. In the exterior region we observe Friedel oscillations that arise from the scattering of Dirac quasiparticles. In contrast to semiconductor quantum dots these graphene quantum dots are fully exposed and directly accessible to real-space imaging tools. The techniques presented here might be extended to more complicated systems such as multiple quantum dots[31-33] with variable coupling and arbitrary geometries.

**Acknowledgments:**

The authors thank A.N. Pasupathy (who has similar work by Gutiérrez *et al.*), J.A. Stroscio, N.B. Zhitenev, and J. Wyrick for stimulating discussions. This research was supported by the sp$^2$ program (STM measurement and instrumentation) and the LBNL Molecular Foundry (graphene characterization) funded by the Director, Office of Science, Office of Basic Energy Sciences of the US Department of Energy under Contract No. DE-AC02-05CH11231. Support was also provided by National Science Foundation award CMMI-1206512 (device fabrication, image analysis). D.W. was supported by the Department of Defense (DoD) through the National Defense Science & Engineering Graduate Fellowship (NDSEG) Program, 32 CFR 168a.

**Author Contributions:**

J.V.J., D.W., and J.L. conceived the work and designed the research strategy. D.W., J.L., J.F.R.N., and J.V.J. performed data analysis. S.K., J.V.J., and A.Z. facilitated sample fabrication. J.L., D.W., and J.V.J. carried out STM/STS measurements. K.W. and T.T. synthesized the h-BN samples. J.F.R.N. and L.S.L performed theoretical calculations. M.F.C. supervised the STM/STS experiments. J.L., D.W., J.V.J, J.F.R.N., L.S.L., and M.F.C. co-wrote the manuscript. D.W. and M.F.C. coordinated the collaboration. All authors discussed the results and commented on the paper.

## Methods:

**Sample Fabrication.** We fabricated our samples using a transfer technique developed by Zomer et al.[34] that uses 60-100 nm thick BN crystals (synthesized by Taniguchi and Watanabe[35]) and 300 nm thick $SiO_2$ as the dielectric for electrostatic gating. Single-layer graphene was mechanically exfoliated from graphite and deposited onto methyl methacrylate (MMA) before being transferred onto BN previously exfoliated onto a heavily doped $SiO_2$/Si wafer. The graphene was electrically grounded through a Ti (10 nm)/Au (40-100 nm) electrode deposited via electron beam evaporation using a shadow mask. Devices were annealed in Ar/$H_2$ gas at 350°C before being transferred into our Omicron ultra-high vacuum (UHV) low temperature STM. A second anneal was performed overnight at 250-400°C and $10^{-11}$ torr.

**Scanning tunneling microscopy and spectroscopy measurements.** All STM measurements were performed at T = 4.8 K with a platinum iridium STM tip calibrated against the Au(111) Shockley surface state. STM topographic and $dI/dV_s$ images were obtained at constant current with sample bias $V_s$ defined as the negative of the voltage applied to the STM tip with respect to the grounded graphene sample. Scanning tunneling spectroscopy (STS) measurements were performed by lock-in detection of the a.c. tunnel current induced by a modulated voltage (1-6 mV at 613.7 Hz) added to $V_s$, while $dI/dV_s(V_g,V_s)$ and $dI/dV_s(r,V_s)$ measurements were acquired by sweeping $V_s$ (starting from a fixed set of initial tunneling parameters) and then incrementing $V_g$ or $r$ ($dI/dV_s(V_g,V_s)$ and $dI/dV_s(r,V_s)$ measurements were restricted to -0.1 eV < $V_s$ < 0.1 eV to avoid energy broadening induced by phonon-assisted inelastic tunneling[36]). All $d^2I/dV_s^2$ figures are numerical derivatives of $dI/dV_s$ with respect to $V_s$. These results were reproduced with numerous STM tips on more than 30 pn junction structures.

**Theoretical modeling.** The eigenstates of the Dirac equation are obtained by solving $[v_F \boldsymbol{\sigma} \cdot \boldsymbol{p} + U(\boldsymbol{r})] \Psi(\boldsymbol{r}) = \varepsilon \Psi(\boldsymbol{r})$, where $U(\boldsymbol{r})$ is the electrostatic potential and $\boldsymbol{p} = -i\hbar \nabla_r$. Because $U(\boldsymbol{r})$ is radially symmetric, we use the polar decomposition ansatz

$$\Psi_m(r, \theta) = \frac{e^{im\theta}}{\sqrt{r}} \begin{pmatrix} u_1(r) e^{-i\theta/2} \\ iu_2(r) e^{i\theta/2} \end{pmatrix}$$

with $m$ a half-integer. By inserting the ansatz into the eigenvalue equation we obtain

$$\begin{pmatrix} [U(r) - \varepsilon]/\hbar v_F & \partial_r + m/r \\ -\partial_r + m/r & [U(r) - \varepsilon]/\hbar v_F \end{pmatrix} u(r) = 0$$

To make direct connection with the STS measurements we calculate the local density of states LDOS($\varepsilon$) as a function of $r$. The LDOS can be written as the sum of $m$-state contributions, LDOS($\varepsilon$) = $\sum_m$ LDOS$_m(\varepsilon)$, with

$$\text{LDOS}_m(\varepsilon) = \sum_\nu \langle |u_\nu(r)|^2 \rangle_\lambda \delta(\varepsilon - \varepsilon_\nu)$$

where $\nu$ labels the radial eigenstates for fixed $m$ and $\langle |u_\nu(r)|^2 \rangle_\lambda = \int_0^\infty dr' |u_\nu(r')|^2 e^{-(r-r')^2/2\lambda}$ represents a spatial average of the wavefunction centered at $r$ with a Gaussian weight $\lambda/r^* = 0.01$. To solve the radial Dirac equation, we use the finite difference method discretized in 1200 lattice sites in the interval $0 < r < L$, with a large repulsive potential at $r = L = 12r^*$. Spurious states arising from the finite potential jumps at the boundaries, localized within a few lattice sites of $r = 0$ and $r = L$, are excluded. We sum over eigenstats with azimuthal quantum numbers $-401/2 \le m \le 401/2$, which is sufficient to accurately represent the states in the energy range of interest. The delta function is approximated as a Lorentzian with width $0.3\varepsilon^*$.

**Figure 1 | Creating and imaging a circular graphene pn junction. a,** Schematic diagram showing the fabrication of a local embedded gate in a graphene/BN heterostructure. A square voltage pulse is applied to the STM tip (held a few nanometers from the graphene surface) while the backgate voltage $V_g$ is fixed at a nonzero value. $V_s$ is defined as the negative of the tip bias. This technique creates a circular pn junction in the graphene in response to trapped space charge in the insulating BN. **b,** The STM tip spatially probes Dirac fermion wavefunctions in the presence of the pn junction. **c,** A representative experimental charge density map for one quadrant of a circular graphene pn junction. A $dI/dV_s$ spectrum is measured at each pixel to determine the Dirac point energy $E_D(x,y)$, which is then converted to a local charge carrier density $n(x,y)$. The black dashed line marks the approximate location of the pn junction boundary at $V_g = 40$ V.

**Figure 2 | Gate-tunable electronic structure of a circular graphene pn junction. a,** Schematic diagram of a circular pn junction formed in graphene. The white rectangle indicates the measurement region. **b,** STM topographic image of region sketched in **a**. **c,** $dI/dV_s$ map of same region shown in **b**. Dashed lines are placed near the pn junction boundary in **b** and **c** to serve as guides to the eye. ($V_s = -0.25$ V, $I = 0.5$ nA, $V_g = 30$ V, 6 mV rms a.c. modulation added to $V_s$). **d-g,** $d^2I/dV_s^2(V_g, V_s)$ plots measured at different distances from the center as indicated in **c** (initial tunneling parameters: $V_s = -0.1$ V, $I = 1.5$ nA, 1 mV a.c. modulation). The observed resonances vary in energy roughly according to the expected graphene dispersion $\varepsilon \propto \sqrt{|V_g - V_{CNP}|}$. The energy spacing between resonances is larger at the center (**d**) than seen further out (**f**), and the resonances disappear altogether beyond the pn junction boundary (**g**).

**Figure 3 | d$I$/d$V_s$ images of quantum interference throughout a circular graphene pn junction. a,** d$I$/d$V_s$ map measured for a pn junction sector similar to Fig. 2a but having opposite heterojunction polarity ($V_s$ = 25 mV, $I$ = 0.5 nA, $V_g$ = -23 V, 1 mV a.c. modulation). **b,** d$I$/d$V_s$ map at the same location as **a** but for a different energy shows a different spatial pattern ($V_s$ = 22 mV, $I$ = 0.4 nA, $V_g$ = -22 V, 1 mV a.c. modulation). The dark bands (low d$I$/d$V_s$) marked by the dashed lines in the middle of **a** and **b** represent the classical turning points of the potential.

**Figure 4 | Spatially resolving energy levels inside a circular graphene pn junction. a,** d$^2I$/d$V_s^2$ measured as a function of $V_s$ and radial distance $r$ from the center of a circular pn junction having same polarity as Fig. 2a. The measurement was performed at a fixed gate voltage (initial tunneling parameters: $V_g$ = 32 V, $V_s$ = -0.1 V, $I$ = 1.5 nA, 1 mV a.c. modulation). **b,** Theoretically simulated $\partial$LDOS/$\partial\varepsilon$ as a function of energy and radial distance for a potential $U(r) = -\kappa r^2$ (potential shown as dashed line). **c,** Experimental d$I$/d$V_s$ radial line scans at different $V_s$ values for fixed $V_g$ = 32 V. **d,** Radial dependence of the theoretical probability density ($|\Psi_{n,m}|^2$) for quantum dot eigenstates. Each curve is labeled by radial and azimuthal quantum numbers $(n, m)$. Each set of theoretical curves has been vertically displaced by a quantity proportional to $V_s$ for the correspondingly colored experimental curve in **c** to ensure that the black dashed line denotes the classical turning points.

# References:

1. Katsnelson, M.I., Novoselov, K.S. & Geim, A.K. Chiral tunnelling and the Klein paradox in graphene. *Nat Phys* **2**, 620-625 (2006).
2. Young, A.F. & Kim, P. Quantum interference and Klein tunnelling in graphene heterojunctions. *Nat Phys* **5**, 222-226 (2009).
3. Stander, N., Huard, B. & Goldhaber-Gordon, D. Evidence for Klein Tunneling in Graphene p-n Junctions. *Physical Review Letters* **102**, 026807 (2009).
4. Cheianov, V.V. & Fal'ko, V.I. Selective transmission of Dirac electrons and ballistic magnetoresistance of n-p junctions in graphene. *Physical Review B* **74**, 041403 (2006).
5. Shytov, A.V., Rudner, M.S. & Levitov, L.S. Klein Backscattering and Fabry-Pérot Interference in Graphene Heterojunctions. *Physical Review Letters* **101**, 156804 (2008).
6. Velasco Jr., J. *et al.* Nanoscale control of rewriteable doping patterns in pristine graphene/boron nitride heterostructures. *Nano Letters* **16**, 1620-1625 (2016).
7. Ponomarenko, L.A. *et al.* Chaotic Dirac Billiard in Graphene Quantum Dots. *Science* **320**, 356-358 (2008).
8. Schnez, S. *et al.* Imaging localized states in graphene nanostructures. *Physical Review B* **82**, 165445 (2010).
9. Todd, K., Chou, H.-T., Amasha, S. & Goldhaber-Gordon, D. Quantum Dot Behavior in Graphene Nanoconstrictions. *Nano Letters* **9**, 416-421 (2009).
10. Han, M.Y., Özyilmaz, B., Zhang, Y. & Kim, P. Energy Band-Gap Engineering of Graphene Nanoribbons. *Physical Review Letters* **98**, 206805 (2007).
11. Allen, M.T. *et al.* Spatially resolved edge currents and guided-wave electronic states in graphene. *Nat Phys* **advance online publication**(2015).
12. Subramaniam, D. *et al.* Wave-Function Mapping of Graphene Quantum Dots with Soft Confinement. *Physical Review Letters* **108**, 046801 (2012).
13. Hämäläinen, S.K. *et al.* Quantum-Confined Electronic States in Atomically Well-Defined Graphene Nanostructures. *Physical Review Letters* **107**, 236803 (2011).
14. Phark, S.-h. *et al.* Direct Observation of Electron Confinement in Epitaxial Graphene Nanoislands. *ACS Nano* **5**, 8162-8166 (2011).
15. Lu, J., Yeo, P.S.E., Gan, C.K., Wu, P. & Loh, K.P. Transforming $C_{60}$ molecules into graphene quantum dots. *Nat Nano* **6**, 247-252 (2011).
16. Jung, S. *et al.* Evolution of microscopic localization in graphene in a magnetic field from scattering resonances to quantum dots. *Nat Phys* **7**, 245-251 (2011).
17. Wang, Y. *et al.* Observing Atomic Collapse Resonances in Artificial Nuclei on Graphene. *Science* **340**, 734-737 (2013).
18. Downing, C.A., Stone, D.A. & Portnoi, M.E. Zero-energy states in graphene quantum dots and rings. *Physical Review B* **84**, 155437 (2011).
19. Wu, J.-S. & Fogler, M.M. Scattering of two-dimensional massless Dirac electrons by a circular potential barrier. *Physical Review B* **90**, 235402 (2014).
20. Schulz, C., Heinisch, R.L. & Fehske, H. Scattering of two-dimensional Dirac fermions on gate-defined oscillating quantum dots. *Physical Review B* **91**, 045130 (2015).

21. Chen, H.-Y., Apalkov, V. & Chakraborty, T. Fock-Darwin States of Dirac Electrons in Graphene-Based Artificial Atoms. *Physical Review Letters* **98**, 186803 (2007).
22. Matulis, A. & Peeters, F.M. Quasibound states of quantum dots in single and bilayer graphene. *Physical Review B* **77**, 115423 (2008).
23. Bardarson, J.H., Titov, M. & Brouwer, P.W. Electrostatic Confinement of Electrons in an Integrable Graphene Quantum Dot. *Physical Review Letters* **102**, 226803 (2009).
24. Zhao, Y. *et al.* Creating and probing electron whispering-gallery modes in graphene. *Science* **348**, 672-675 (2015).
25. Wong, D. *et al.* Characterization and manipulation of individual defects in insulating hexagonal boron nitride using scanning tunnelling microscopy. *Nat Nano* **10**, 949-953 (2015).
26. Zhang, Y., Brar, V.W., Girit, C., Zettl, A. & Crommie, M.F. Origin of spatial charge inhomogeneity in graphene. *Nat Phys* **5**, 722-726 (2009).
27. Deshpande, A., Bao, W., Miao, F., Lau, C.N. & LeRoy, B.J. Spatially resolved spectroscopy of monolayer graphene on $SiO_2$. *Physical Review B* **79**, 205411 (2009).
28. Xue, J. *et al.* Scanning tunnelling microscopy and spectroscopy of ultra-flat graphene on hexagonal boron nitride. *Nat. Mater.* **10**, 282-285 (2011).
29. Decker, R. *et al.* Local Electronic Properties of Graphene on a BN Substrate via Scanning Tunneling Microscopy. *Nano Lett.* **11**, 2291-2295 (2011).
30. Crommie, M.F., Lutz, C.P. & Eigler, D.M. Confinement of Electrons to Quantum Corrals on a Metal Surface. *Science* **262**, 218-220 (1993).
31. Petta, J.R. *et al.* Coherent Manipulation of Coupled Electron Spins in Semiconductor Quantum Dots. *Science* **309**, 2180-2184 (2005).
32. Fujisawa, T. *et al.* Spontaneous Emission Spectrum in Double Quantum Dot Devices. *Science* **282**, 932-935 (1998).
33. van der Wiel, W.G. *et al.* Electron transport through double quantum dots. *Reviews of Modern Physics* **75**, 1-22 (2002).
34. Zomer, P.J., Dash, S.P., Tombros, N. & van Wees, B.J. A transfer technique for high mobility graphene devices on commercially available hexagonal boron nitride. *Applied Physics Letters* **99**, 232104-232107 (2011).
35. Watanabe, K., Taniguchi, T. & Kanda, H. Direct-bandgap properties and evidence for ultraviolet lasing of hexagonal boron nitride single crystal. *Nature Mater.* **3**, 404-409 (2004).
36. Brar, V.W. *et al.* Observation of Carrier-Density-Dependent Many-Body Effects in Graphene via Tunneling Spectroscopy. *Physical Review Letters* **104**, 036805 (2010).

**FIGURE 1**

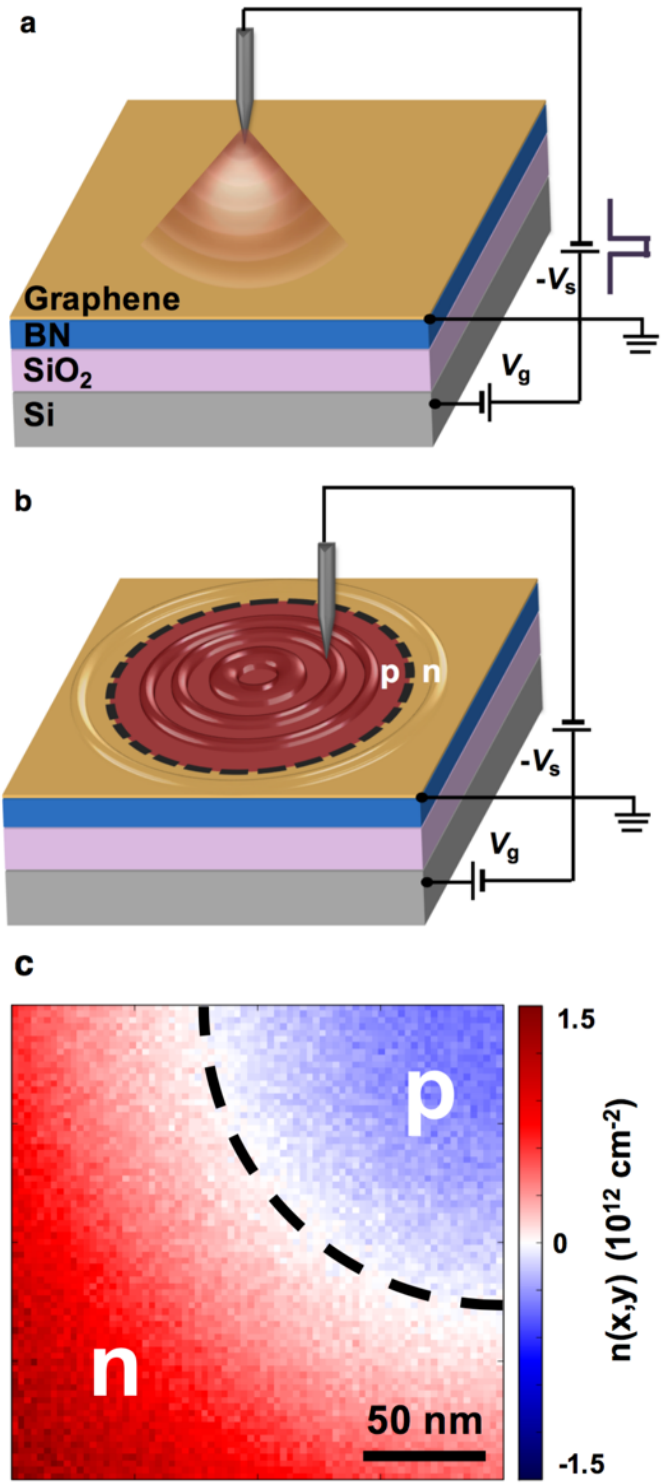

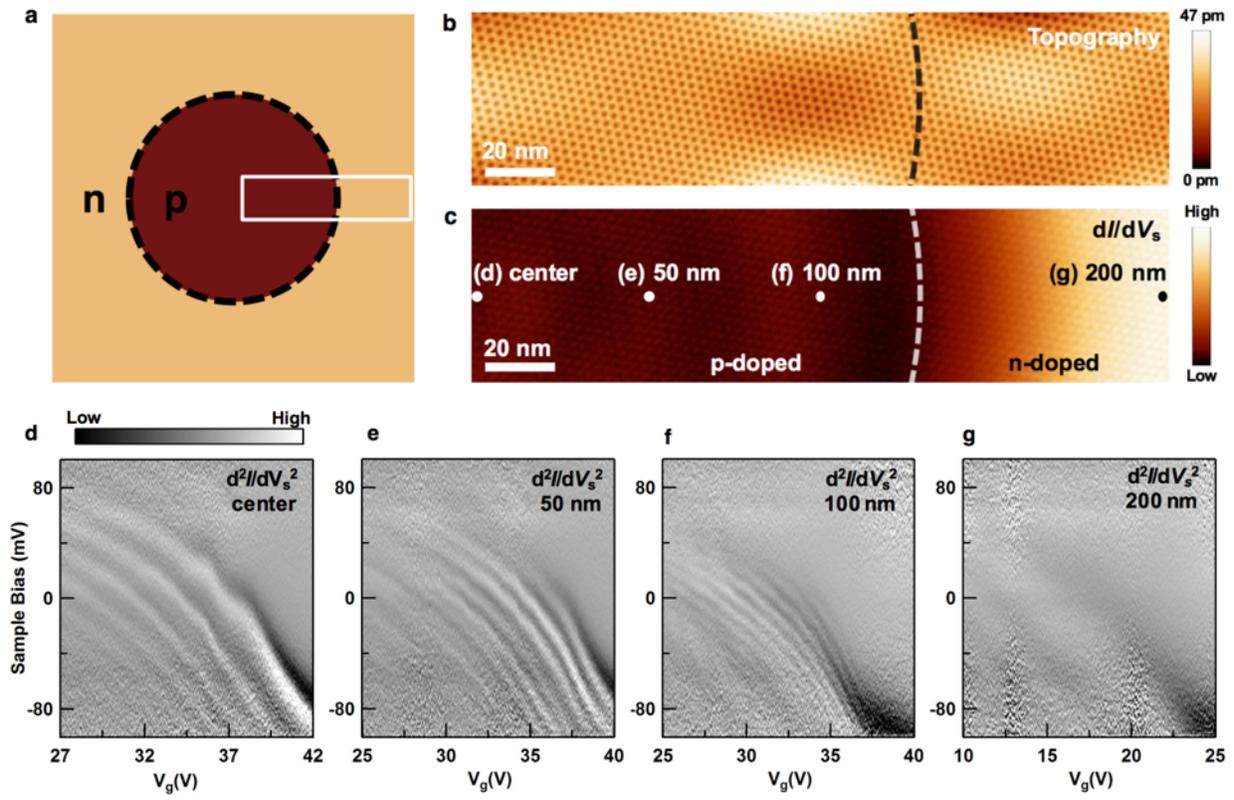

**FIGURE 3**

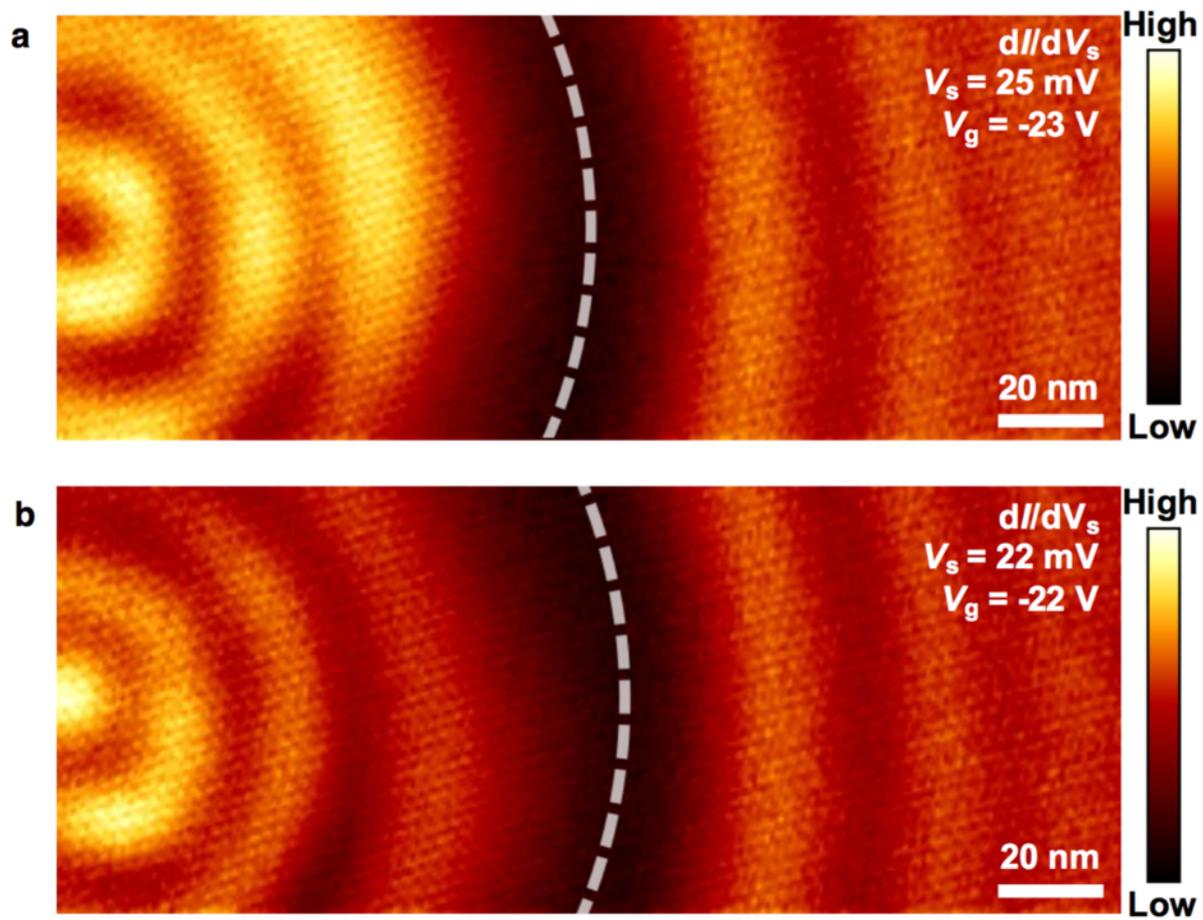

**FIGURE 4**

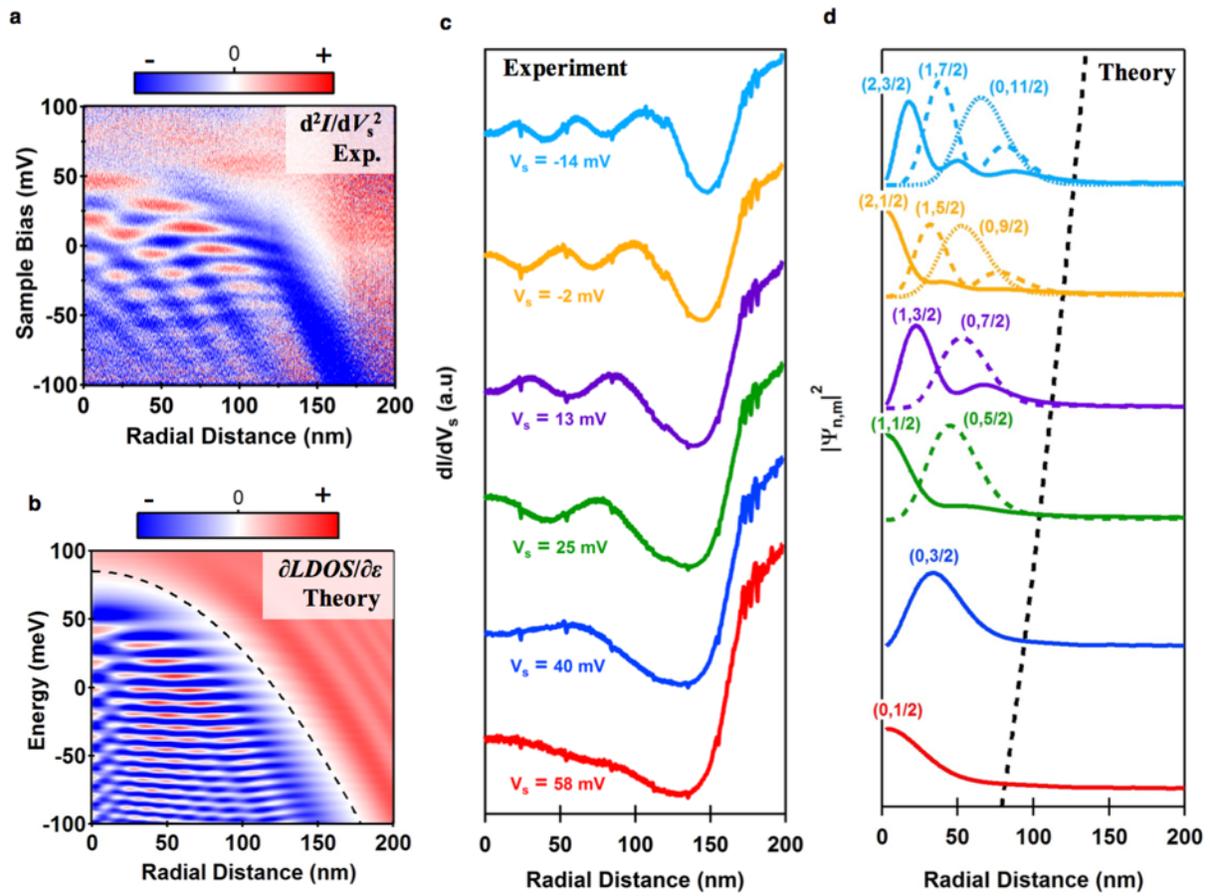